\documentclass[preprint,prd,tightenlines,floatfix,
nofootinbib,eqsecnum,superscriptaddress]{revtex4}

\usepackage[T1]{fontenc}		

\usepackage{soul}
\usepackage{xspace,bm,bbm,dsfont}

\usepackage{amsmath,amsfonts,amssymb,amstext,mathrsfs}
\usepackage{amsthm}
\usepackage{mathpazo}

\usepackage[dvips]{graphicx}
\usepackage{epsf,float}
\usepackage{revsymb}

\usepackage{dcolumn}
\usepackage{braket}
\usepackage{color,xcolor}
\usepackage{graphicx}
\usepackage{subfigure}
\usepackage{multirow}
\usepackage{tabularx}
\usepackage{pstricks}
\usepackage[section]{placeins}
\usepackage{booktabs}
\usepackage{array}

\usepackage{hyperref}

\usepackage{slashed}
\usepackage{multirow}

\usepackage[normalem]{ulem}

\begin{document}


\author{Antoni Szczurek
\footnote{Also at \textit{College of Natural Sciences, 
Institute of Physics, University of Rzesz\'ow, 
Pigonia 1, PL-35310 Rzesz\'ow, Poland}.}}
\email{Antoni.Szczurek@ifj.edu.pl}
\affiliation{Institute of Nuclear Physics Polish Academy of Sciences, 
Radzikowskiego 152, PL-31342 Krak\'ow, Poland}

\title{Local energy-momentum conservation \\ 
and initial conditions for quark-gluon plasma evolution \\
in nucleus-nucleus collisions at SPS and RHIC BES energies}

\begin{abstract}
We investigate consequences of local energy-momentum conservation 
for initial eccentricity coefficients in heavy ion collisions at 
not too high energies relevant for CERN SPS and RHIC BES.
Different models of energy density available for pion production
are considered.
We study dependence of eccentricity coefficients on space-time rapidity 
for different impact parameters.
The naive formula how to define eccentricities breaks with our initial
conditions and must be corrected.
We predict considerable eccentricities for $\epsilon_{1,2,3,4}$
and specific dependence on space-time rapidity as well as on impact
parameter for lower energies.
The effect becomes smaller at larger energies when restricting to 
narrow rapidity interval arround zero.
Our predictions are in principle input for further hydrodynamical
evolution but it is not clear whether they can be easily used.
Our initial condition suggest a strong preequilibrium phase
which is difficult for modelling.
\end{abstract}

\maketitle

\section{Introduction}

The initial conditions for hydrodynamical evolution of quark-gluon
plasma is of crucial importance for final observables such as rapidity
distributions of the produced pions or flow parameters and their
dependence of particle (pseudo)rapidity and collision centrality.
So far mostly Glauber-collision or color-glass-condensate initial
conditions were most often used in this context 
\cite{Glissando,SS2016,LKE2018,SS2018}.
They lead to more-or-less correct behaviour of the flow at midrapidities
which is of special interest for plasma studies.

In Ref.\cite{SKR2017} we suggested that the conservation
of energy and momentum in local longitudinal streaks, dependent on the position
in the transverse momentum plane (x,y), provides strong constraints on
initial conditions for plasma evolution in heavy ion collisions.
In this model the produced plasma in peripheral collisions is not at rest
and its movement depends on the position in the impact parameter
space $\vec{\rho} = (x,y)$. In particular, the parts of plasma that 
are close to spectators move with velocities only slightly smaller than
velocities of spectators while those for $\vec{\rho}(0,0)$ are at rest
on average. Such initial conditions were used recently in studies of
electromagnetic effects caused by strong fields generated by fast moving
spectators and a good description of its Feynman-$x_F$ dependence
was obtained \cite{RS2007} in agreement with the NA49 data 
on the $\pi^+/\pi^-$ ratio at low pion transverse momenta
\cite{ORSMK2020}. Similar idea was studied
recently in Ref.\cite{SA2020} where so-called directed flow of pions 
and protons was studied. The two approaches differ in defining
energy available for particle production which will be discussed here in
detail.

The important ingredient of the modelling
of pion emission in \cite{SKR2017} was fragmentation function, assumed 
universal for different impact parameters.
There the fragmentation function parametrizes both nuclear transparency,
evolution of plasma and fluid-to-hadron transition in an effective way,
so it must be collision energy dependent.
The fragmentation function is then adjusted to experimental data.
In \cite{SA2020} the transparency or intial plasma distribution is
parametrized and the fluid-to-hadron transition is a matter of a
procedure applied after termination of the hydrodynamical evolution,
not explained in extenso in \cite{SA2020}. The latter may be not well
under control for the specific initial conditions for plasma evolution.
It is worth to mention in this context that the production of 
forward/backward hadrons via fragmentation function
is not well understood even for proton-proton collisions \cite{MS2019}.
It becomes even more difficult for heavy hadron production \cite{S2020}.

Here we wish to investigate how the local (in (x,y)) 
energy-momentum conservation as proposed in \cite{SKR2017} 
influences/determines the initial eccentricity parameters of plasma 
created during a peripheral collision 
as a function of space-time rapidity.

\section{Local energy-momentum conservation,
energy available for particle emission
and eccentricities of the produced plasma}

For peripheral collisions of identical nuclei 
(Pb + Pb (SPS) or Au + Au (RHIC)) 
the full stop of nuclear matter is not always possible and 
energy-momentum conservation implies some local collective 
longitudinal flow of parts of the plasma.
As discussed in \cite{SKR2017} the tips of the almond-like initial 
fire-ball move very forward or very backward with large velocities, 
so the inital almond-like shape changes quickly with time and after 
a while does not remain the almond shape. Such a picture, very different 
from previously used initial conditions (Glauber or KLN)
must have consequences for rapidity-dependent observables such as
energy dependence on rapidity and Fourier coefficients of azimuthal
correlations.

Let us consider for a moment full stopping of nuclear matter i.e.
no nuclear transparency. The full stopping for peripheral collisions
does not mean that the plasma is at rest.
The rapidity of the fully stopped plasma at the ($x,y$) point for the
collision at impact parameter $b$ can be calculated as
\begin{equation}
y_{stop}(x,y;b) = 2 arctanh
\left(\frac{T_A(x,y;b) - T_B(x,y;b)}{T_A(x,y;b) + T_B(x,y;b)} 
tanh(y_{beam}) \right)
\; .
\label{y_stop}
\end{equation}
The so-called thickness functions $T_A$ and $T_B$ are calculated here
based on nucleon (proton and neutron) distributions obtained from
Hartree-Fock-Bogoliubov method \cite{M2000} which includes neutron skin 
effects.
$y_{stop}$ above can be interpreted as rapidity of the plasma contracted
to a point in $z$ direction, of course different for different ($x,y$).
This is a text-book example of fully inelastic collision.
In this approach the plasma moves with different rapidities for different
position in the $(x,y)$ space and different impact parameter $b$.

The distribution of plasma in space-time rapidity with respect 
to $y_{stop}(x,y; b)$, rapidity of the fully stopped (no nuclear
transparency) plasma at the point $(x,y)$
cannot be caluclated at present and must be parametrized.
\footnote{There are first trials to calculate baryon stopping,
see e.g. \cite{LSS2019}. The relevant experimental data were obtained
some time ago by e.g. the BRAHMS collaboration at RHIC \cite{BRAHMS}.}

As a first option we consider a Gaussian parametrization of the
plasma-like not equilibrated matter:
\begin{equation}
f(\eta_s;x,y,b) = \frac{1}{\sqrt{2 \pi \sigma_{\eta}^2}}
\exp \left( \frac{(\eta_s - y_{stop}(x,y;b))^2}{2 \sigma_{\eta}^2}
\right) \; .
\label{f_etas_Gaussian}
\end{equation}
A rectangular step-like function with the width $d_{\eta}$ could be another
option:
\begin{equation}
f(\eta_s; x,y,b) = const = 1/d_{\eta}
\label{f_etas_step-like}
\end{equation}
for   $\eta_s > y_{stop}(x,y;b) - d_{\eta}/2$ and
      $\eta_s < y_{stop}(x,y;b) + d_{\eta}/2$  and zero otherwise.
The $d_{\eta}$ is a free parameter for the second parametrization.
For simplicity we have assumed universal ((x,y;b)-independent
functions centered at $y_{stop}(x,y; b)$.
At even higher energies (RHIC) a double-Gaussian shape can be considered
\cite{Tomasik}.

In the following we shall call $f(\eta_s; x, y, b)$ ``transparency
function'' for brevity, in spite of the fact that this name may be
slightly confusing, and in order to distinguish from 
``fragmentation function'', also somewhat confusing, 
used in \cite{SKR2017},

The authors of Ref. \cite{SA2020} proposed that
the (total) energy density distribution for a given $(x,y)$ transverse 
position point can be calculated as:
\begin{equation}
E(x,y; b) = ( T_A(x,y; b) + T_B(x,y; b) ) \; m_N cosh(y_{beam}) |_{\cal C}
\; .
\label{energy_xy}
\end{equation}
Please note the letter ${\cal C}$ in the formula above which means an extra
condition which must be imposed to eliminate spectator matter from 
the participant matter 
\footnote{It is not clear to us whether it was included in \cite{SA2020}.}.
The extra collision condition ${\cal C}$ written above can be formulated
e.g. as:
\begin{equation}
M_A(x,y) > m_0  \;\;  \text{and} \;\; M_B(x,y) > m_0   \; ,
\label{extra_condition}
\end{equation}
where $M_A$ and $M_B$ are cold mass densities in nucleus $A$ and $B$,
respectively and $m_0$ is a minimal (mass) density required for the
creation of plasma
(a free parameter of the model here).
Such a condition assures that there must be some minimal nuclear matter
portion in both nuclei to create plasma.
Ignoring such a condition would mean ``including spectators'' as
emitters of particles. A precise formulation of the collision condition
${\cal C}$ is not easy close to spectators where the division into
spectator and participant zone takes place.
This condition has sizeable effect on initial eccentricities
as will be discussed in the following.

For comparison the cold mass taking part in the collision 
can be written as:
\begin{equation}
M(x,y; b) = ( T_A(x,y; b) + T_B(x,y; b) ) \; m_N |_{\cal C} \; .
\end{equation}

How the energy is distributed in space-time rapidity requires extra
modelling.
The extra distribution in space-time rapidity can be formally written as 
\begin{equation}
\frac{d E}{d \eta_s} = f(\eta_s;x,y,b) E(x,y) \;
\end{equation}
using the phenomenological functions 
$f(\eta_s; x,y,b) = f(\eta_s; y_{stop})$ (Gaussian, step-like) 
discussed above.

The space-time rapidity distribution of energy contained in plasma 
for a given impact parameter $b$ can be calulated as:
\begin{equation}
\frac{d E}{d \eta_s}\left( \eta_s;b \right)
 = \int \frac{d E}{d \eta_s d x d y}(x,y; \eta_s, b)
\; dx dy  \; . 
\label{E_etas_distribution}
\end{equation}
In our opinion not the whole energy density as is written above
is available for particle production. 

In \cite{SKR2017} we suggested to subtract from the total energy density
$E(x,y; b)$ at least the cold mass density $M(x,y; b)$. 
The cold mass must reappear in the form of produced baryons due to 
baryon-number conservation so corresponding energy cannot be used 
to production of mesons (mostly pions).
Such a correction obviously depends on collision energy and is larger
percentage-wise for lowest energies where plasma is produced. 
Is this correction enough ?

A sizeable part of the energy which is not available to particle production 
is a kinetic energy of parts of plasma moving in forward 
($\eta_s >$ 0) or backward ($\eta_s <$ 0) directions. 
How to calculate the unavailable part of energy to be
subtracted from the total energy for different bins in $x$ and $y$
and a given space-time rapidity $\eta_s$ ($b$ is treated here as a
parameter) ?
A big part of it is contained in final moving(!) baryons. 
Including such a correction is straightforward. The available energy 
can be then obtained as:
\begin{equation}
E_{avail}(x,y;b) = E(x,y;b) - M(x,y;b) \cdot cosh y_{stop}(x,y;b)
|_{\cal C}  > 0 \; .
\label{E_available}
\end{equation}
The $(x,y)$ profile of the newly defined quantity $E_{avail}(x,y;b)$ 
is obviously different than that for total energy density $E(x,y; b)$.
In Ref.\cite{SA2020} the subtraction term above was not included.
It may be crucial which energy is used as initial condition for plasma
hydrodynamical evolution.
We shall discuss consequences of the correction 
(not included in \cite{SA2020}) both on
particle production and on eccentricties in the following.

To illustrate the global situation related to energy available in the
collision or available for the particle production
we define the integrated energy densities:
\begin{eqnarray}
E (b) &=& \int dx dy E(x,y; b) \; , \nonumber \\
E_{avail}(b) &=& \int dx dy E_{avail}(x,y;b) \; .
\label{integrated_energies}
\end{eqnarray}
Very interesting is the energy distribution per space-time rapidity unit
for a given impact parameter (centrality) which is strongly related to 
particle rapidity distribution. Formally it can be defined as:
\begin{eqnarray}
\frac{d E}{d \eta_s}(\eta_s;b) &=& 
\int \frac{d E(x,y;b)}{d \eta_s} \; dx dy
\; , \\
\frac{d E_{avail}}{d \eta_s;b}(\eta_s) &=& 
\int \frac{d E_{avail}(x,y;b)}{d \eta_s} \; dx dy \; .
\label{etas_energy_distribution}
\end{eqnarray}

The eccentricities of produced plasma, defined usually 
for midrapidities, read (see e.g. \cite{NYGO2016}):
\begin{equation}
\epsilon_n = - \frac{ \int exp(+i n \phi) \rho^n E(\rho,\phi) \; d^2 \rho}
                    { \int \rho^n E(\rho,\phi) \; d^2 \rho} \; ,
\label{epsilon_n_definition}
\end{equation}
where $\rho = \sqrt{x^2 + y^2}$ and $\phi$ is azimuthal angle with
respect to the reaction plane.
Is this definition universal and valid also for forward/backward 
space-time rapidities ?
We shall return to the issue in the next section when discussing
energy density as a function of $(x,y,\eta_s)$.

In the next section we shall compare our results for $\epsilon_n$ for 
the two proposed parametrizations of energy distribution in $\eta_s$ 
and different ways of weighting the energy available for meson production.

\section{Numerical results and discussion}

\begin{figure}
\includegraphics[width=7cm]{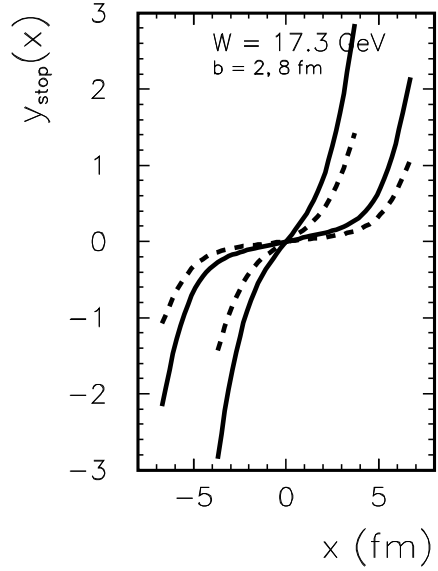}
\caption{
Rapidity of the stopped plasma $y_{stop}$ as a function of $x$ for 
$y$ = 0 and two different values of impact parameter $b =$ 2 (black lines),
8 (red lines) fm.
The solid lines are in agreement with \cite{SKR2017} while the dashed lines
are obtained according to \cite{SA2020}.
}
\label{fig:ystop_x}
\end{figure}

We wish to start our discussion by showing the collective longitudinal motion
of plasma created in the high-energy collision.
In Fig.\ref{fig:ystop_x} we show dependence of rapidity of the stopped plasma
as a function of $x$ for fixed $y$ = 0 for two different values
of impact parameter $b$ = 2, 8 fm. We wish to note the difference
of our result (solid line) compared to that from \cite{SA2020} (dashed
line). We predict that the pieces of plasma that are close to spectators
have almost the same space-time rapidities as spectators.
This is universal feature independent of collision energy.

\begin{figure}
\includegraphics[width=7cm]{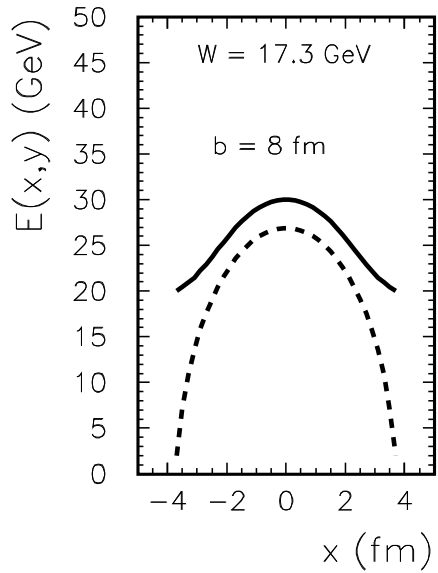}
\caption{
Energy density calculated for the same geometrical situation as
in the previous figure for $b$ = 8 fm collision. The dashed line
represents result obtained from Eq.(\ref{energy_xy}) while the solid line
takes into account the subtraction of the collective longitudinal
energy, which is useless for particle production.
}
\label{fig:energy_density_x}
\end{figure}

The corresponding ($x,y$)-dependent energy density is shown in 
Fig.\ref{fig:energy_density_x}. We show the results obtained using
the naive formula (\ref{energy_xy}) (dashed line) and the energy corrected
for the collective longitudinal energy which cannot be used up
for particle production. There is a significant effect for positions
in $x$ close to spectators. This means a reduced production of particles
(pions) from this region of configuration space. This means in a hidden
way also earlier creation of mesons (pions) from these parts of plasma. 
This must have direct consequences for electromagnetic effect on pions 
generated by fast moving spectators. The earlier produced pions feel 
the EM field of spectators somewhat earlier.

Now we wish to present influence of the longitudinal flow on integrated 
quantities defined in (\ref{integrated_energies}).
We observe that the energy available for particle production (solid line) 
is somewhat smaller, about 10 \%, than the total energy 
in the collision (dashed line).
The relative global effect only weakly depends on impact parameter.
So the effect for minimum bias collisions is rather small.

\begin{figure}
\includegraphics[width=7cm]{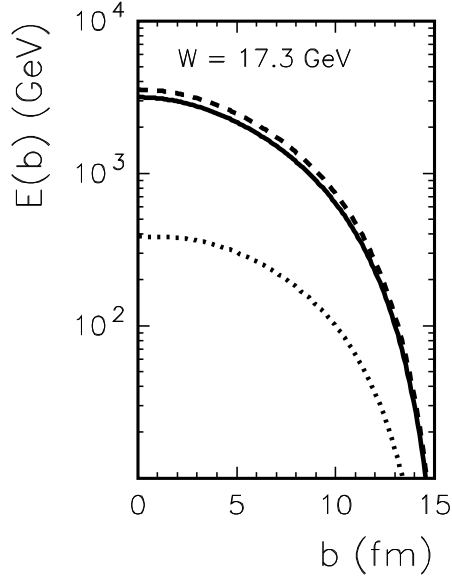}
\caption{
Dependence of integrated energies defined by (\ref{integrated_energies})
on impact parametr for
$\sqrt{s_{NN}}$ = 17.3 GeV. The $E(b)$ is shown as a
the dashed line, while $E_{avail}(b)$ as the solid line
and their difference as the dotted line.
}
\label{fig:integrated_energies_b}
\end{figure}

\begin{figure}
\includegraphics[width=6cm]{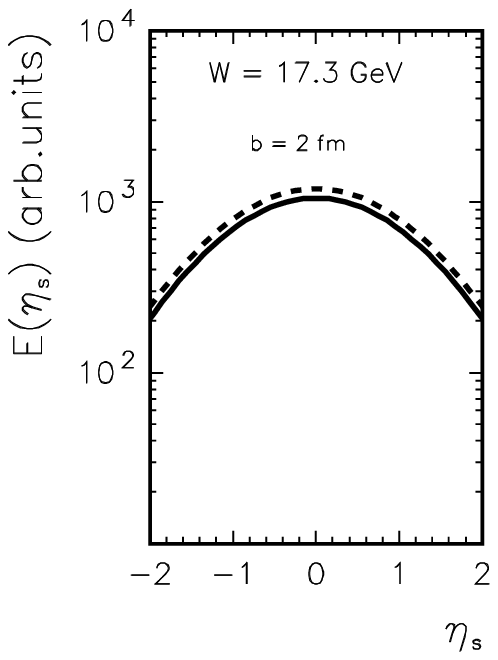}
\includegraphics[width=6cm]{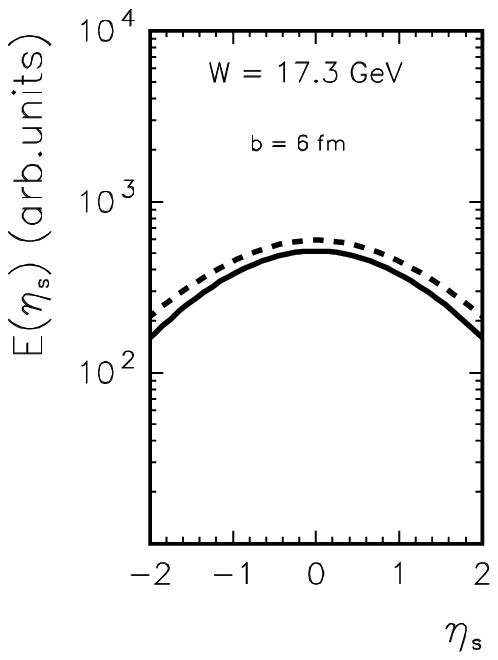}
\caption{
The distribution of total (dashed line) and available (solid line) energy
for $b =$ 2 fm (left panel) and $b =$ 6 fm (right panel).
In this calculation $\sqrt{s_{NN}}$ = 17.3 GeV, $\sigma_{\eta}$ = 1.0.
}
\label{fig:dE_detas}
\end{figure}

The distribution of energy per unit of space-time rapidity is shown
in Fig.\ref{fig:dE_detas} for two different impact parameters 
b = 2 fm (left panel) and b = 8 fm (right panel).
We show distribution of $d E / d \eta_s$ (dashed line) and
$d E_{avail} / d \eta_s$ (solid line).
The shape of the distribution should remind the shape of the rapidity
distribution of pions which is known for $\sqrt{s_{NN}}$ = 17.3 GeV.
As in Ref.\cite{SKR2017} we get slightly broader distribution
in $\eta_s$ for peripheral collisions. The $\sigma_{\eta}$ = 1.0
give more or less realistic distributions. $\sigma_{\eta}$ = 0.3
from \cite{SA2020} gives in our opinion too narrow distribution
and should be increased.

\begin{figure}
\includegraphics[width=5cm]{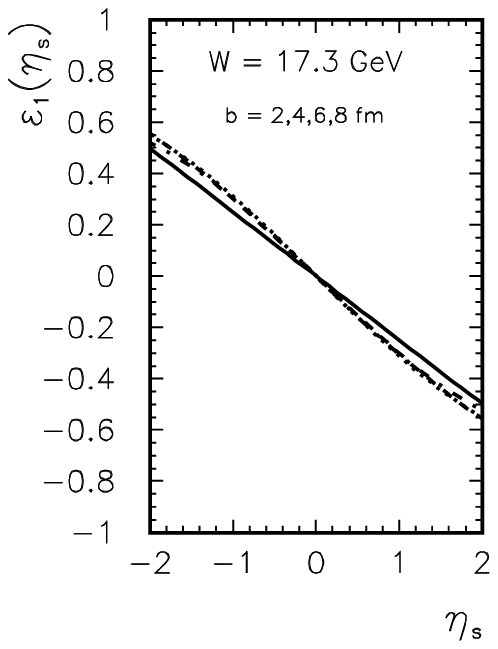}
\includegraphics[width=5cm]{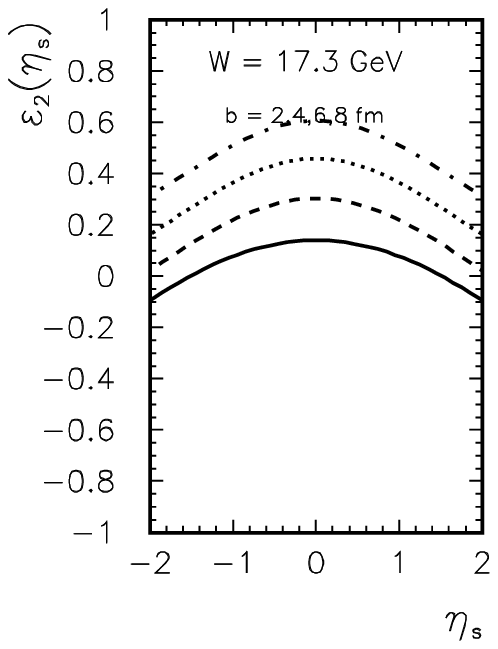} \\
\includegraphics[width=5cm]{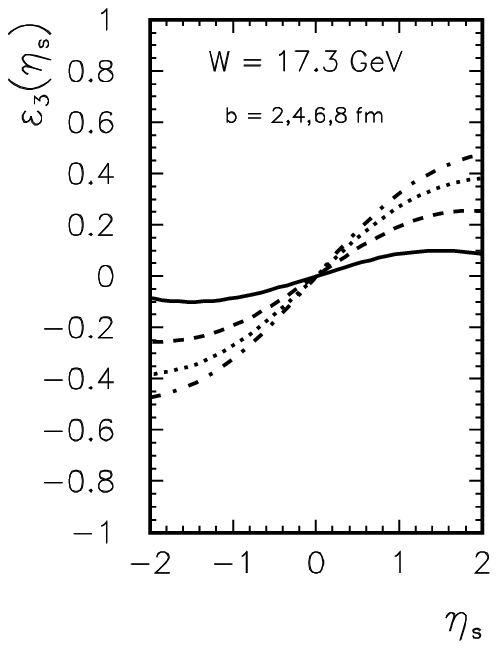}
\includegraphics[width=5cm]{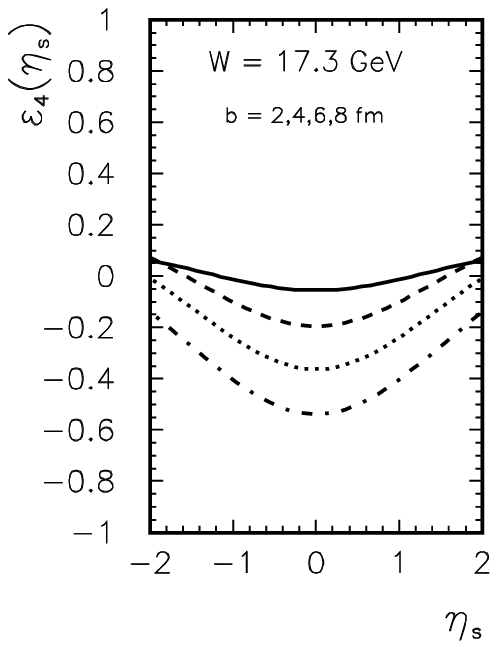}
\caption{Initial eccentricity parameters as a function of
space-time rapidity calculated from
formulae (\ref{epsilon_n_definition}) for $\sqrt{s_{NN}}$ = 17.3 GeV
for different values of impact parameter $b$ = 2, 4, 6, 8 fm.
}
\label{fig:epsilons_naive}
\end{figure}

Now we will proceed to eccentricity coefficients.
In Fig.\ref{fig:epsilons_naive} we collected the dependences
of the lowest eccentricities as a function of space-time rapidity
for different impact parameters for $Pb + Pb$ collisions
at $\sqrt{s_{NN}}$ = 17.3 GeV relevant for the NA49 experiment. 
Here we have used $\sigma_{\eta}$ = 1.0 as adjusted roughly to 
experimental data on pion rapidity distributions.
The eccentricity coefficients calculated according to formula 
(\ref{epsilon_n_definition})
depend strongly on rapidity. The even coefficients depend also
strongly on the impact parameter.
The so-calculated eccentricity coefficients
contain spourious shift of the forward/backward moving matter from 
the (0,0) point.


\begin{figure}
\includegraphics[width=5.2cm]{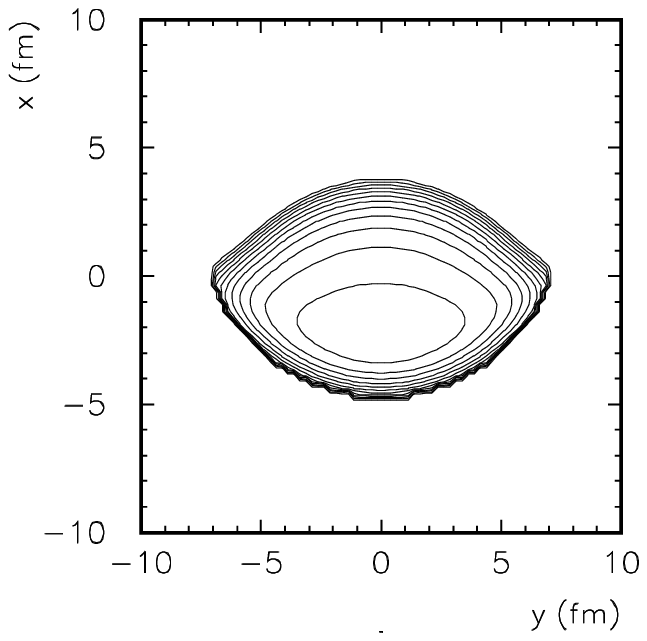}
\includegraphics[width=5.2cm]{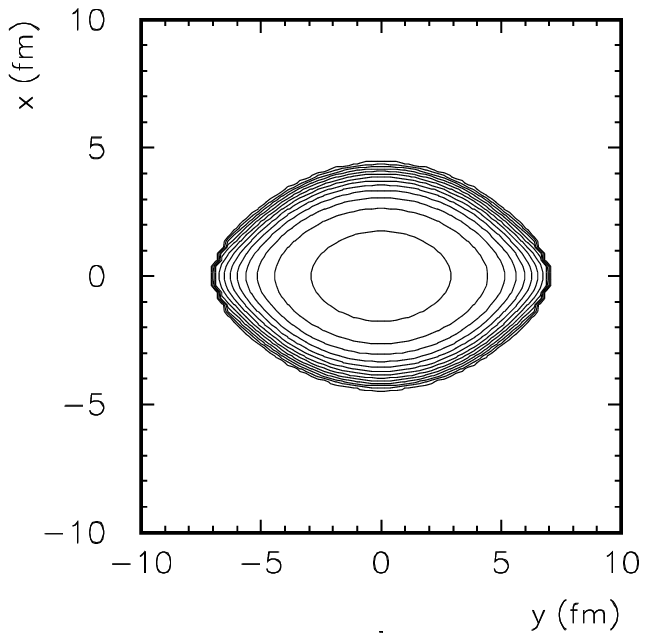}
\includegraphics[width=5.2cm]{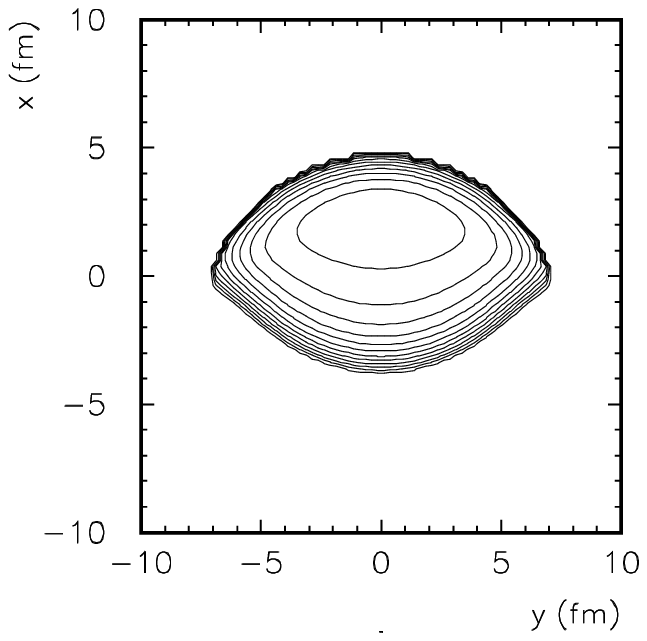}\\
\includegraphics[width=5.2cm]{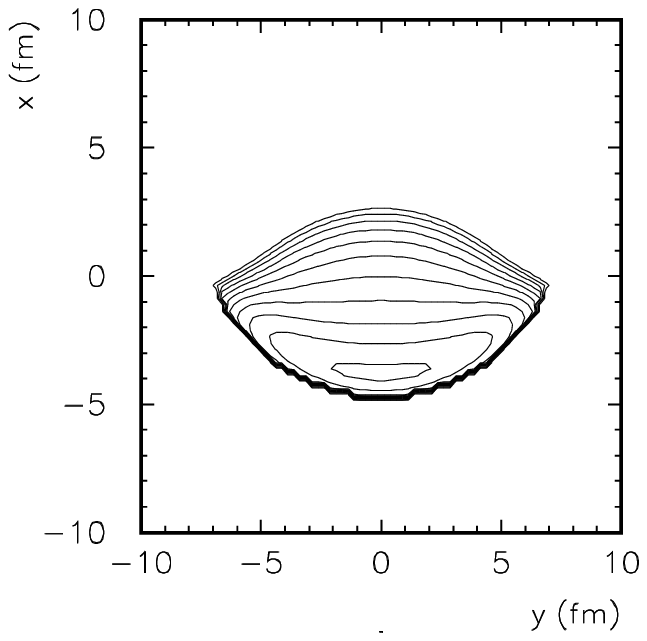}
\includegraphics[width=5.2cm]{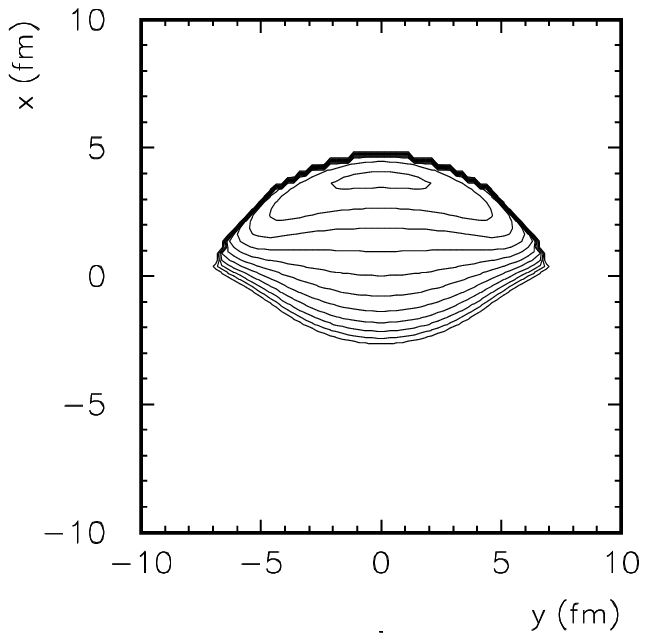}
\caption{Density of energy available for particle production 
as a function of the position in the
transverse space for five different intervals of space-time
rapitity: $\eta_s$ =-1 (upper left panel), 
          $\eta_s$ = 0 (upper middle panel),
      and $\eta_s$ = 1 (upper right panel)
and for $\eta_s$ = $\mp$ 2 (lower panels).
Here $\sqrt{s_{NN}}$ = 17.3 GeV and b = 6 fm and the Gaussian
transparency function was used.
}
\label{fig:energy-density_xy_fixed_etas}
\end{figure}

In Fig.\ref{fig:energy-density_xy_fixed_etas} we present energy
available for particle emission as a function of $(x,y)$ 
for the Gaussian distribution in $\eta_s$ (see Eq.(\ref{f_etas_Gaussian})), 
for fixed values of $\eta_s$ = -2, -1, 0, 1, 2 with the $\eta_s$-window width 
$\pm$ 0.1. 
The shapes for larger $\eta_s$ become more eccentric and naturaly should
have the whole spectrum of eccentricity coefficients.
Pieces of plasma with large $|\eta_s|$ have funny shapes in $(x,y)$.
Equilibration of plasma with such strange shapes cannot be fast, or 
does not occur at all as the remote parts do not communicate between 
themselves due to causality. Here naturally preequilibrium evolution must
be at work.


We clearly see that
x=0, y=0 point is inadequate for defining eccentricity coefficients for 
forward or backward going plasma. In this case one should find more 
adequate reference point for shifted $x$, centered at the middle 
of the plasma at a given space-time rapidity.

In our approach the situations shown in upper left panel and upper right
panel as well as in lower left and lower right panels are correlated by 
the geometry. The parts of plasma going in forward and backward
directions are correlated by geometry in the configuration space.
This correlation translates (via preequilibrium and/or hydro evolution)
to correlation in the momentum space which leads to forward-backward
azimuthal correlations. Such an effect was discussed first in 
\cite{BBM2011} in a different approach. We do not know if this was 
observed experimentally.
This is also related to forward-backward multiplicity correlations
which in our case is strongly related to the energy available 
for particle emission.
Such correlations have been observed experimentally
\cite{FB_correlations_PHOBOS,FB_correlations_STAR}. 
However, in this case fluctuations are crucial.

The standard (hidden) reference point used for midrapidities $x=0, y=0$
(see Eq.(\ref{epsilon_n_definition})) must be 
modified to $x=x_{shift}, y=0$ for nonzero $\eta_s$
 \footnote{We use similar procedure as proposed in \cite{SA2020} but our
 weights seem different.}
\begin{equation}
x_{shift}(\eta_s; b) = \frac{\int E(x,y;\eta_s,b) x \; dx dy}
                        {\int E(x,y;\eta_s,b) \; dx dy} \; .
\label{x_shift}
\end{equation}
Above $x_{shift}$ is weighted by energy density ($E$ or $E_{avail}$) 
at a given space-time rapidity slice. 
Then also $\rho$ and $\phi$ in (\ref{epsilon_n_definition}) must 
be modified as:
\begin{eqnarray}
\rho &\to& \tilde{\rho} = \sqrt{(x-x_{shift})^2 + y^2} \; , \\
\phi &\to& \tilde{\phi} \; ,
\label{further_modifications}
\end{eqnarray}
where $\tilde{\phi}$ is a new azimuthal angle in the shifted frame of
reference placed at $(x_{shift},0)$.
$x_{shift}$ is of course defined with the same weight ($E$ or $E_{avail}$)
as used in the definition of $\epsilon_n$.

We shall consider also another option
trying to subtract the unavoidable local longitudinal flow
as discussed in the previous section.

\begin{figure}
\includegraphics[width=8cm]{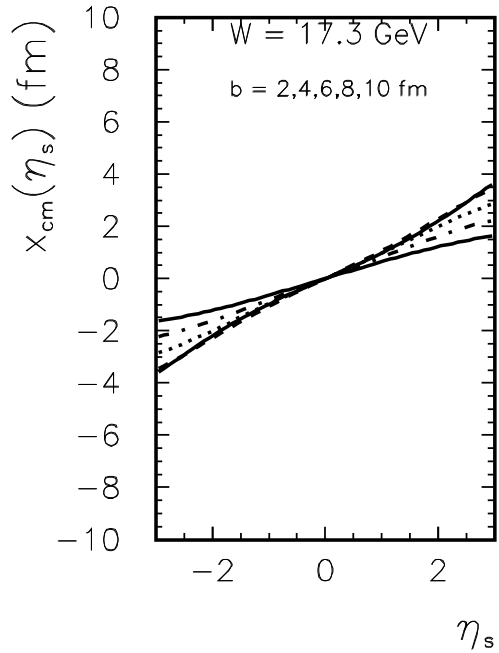}
\caption{$x_{shift}$ defined in (\ref{x_shift}) for $\sqrt{s_{NN}}$ = 17.3 GeV
as a function of $\eta_s$ for five different impact parameters 
$b$ = 2 (solid), 4 (dashed), 6 (dotted), 8 (dash-dotted)
for Gaussian transparency functions with $\sigma_{\eta}$ = 1.
}
\label{fig:x_cm_etas}
\end{figure}

The shift in $x$ is shown in Fig.\ref{fig:x_cm_etas} as a function
of space-time rapidity for different values of the impact parameter. 
The actual dependence is a consequence of the parametrization 
(\ref{f_etas_Gaussian}) 
of energy density as a function of $\eta_s$ which is not known from
first principles.
The shift for large $\eta_s$ is really large, independently of the
parametrization of the transparency function. 
The transparency function related to bayon-stopping mechanism
is therefore crucial for further results. In general, the function
could be adjusted to reproduce rapidity dependent observables.
Some of them will be discussed in the following.

\begin{figure}
\includegraphics[width=5cm]{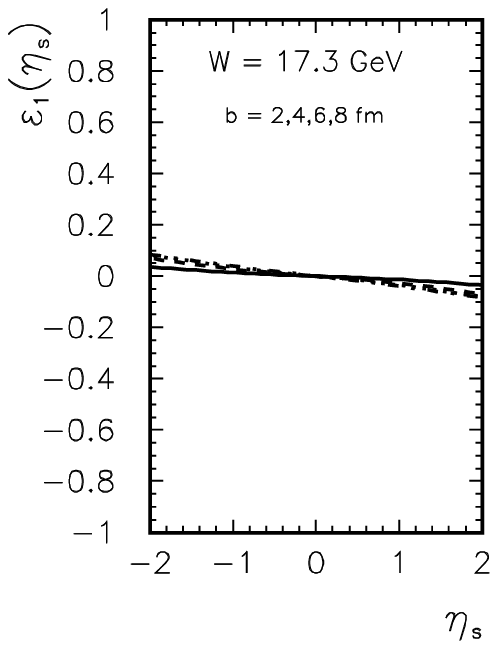}
\includegraphics[width=5cm]{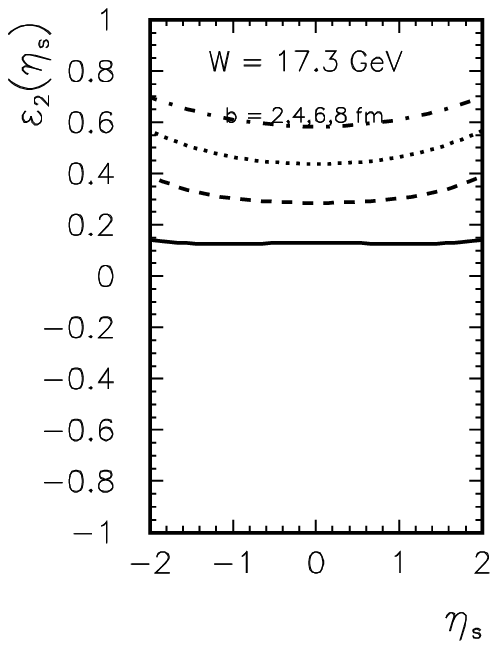} \\
\includegraphics[width=5cm]{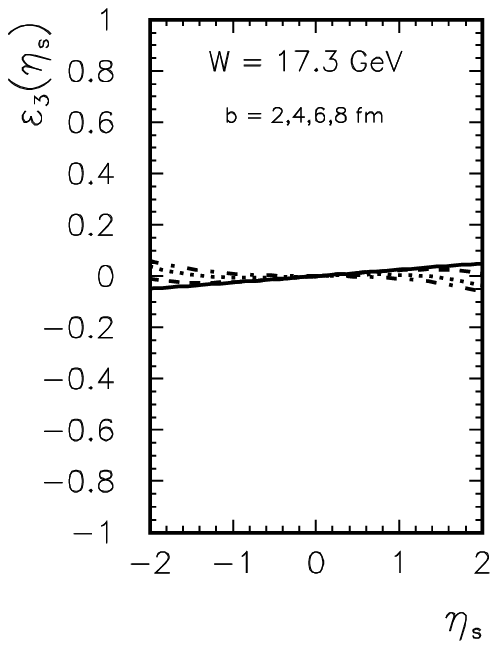}
\includegraphics[width=5cm]{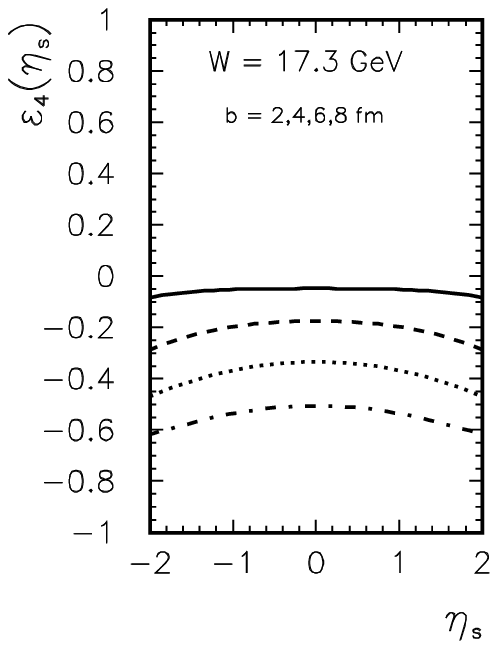}
\caption{Initial eccentricity parameters as a function of
space-time rapidity calculated from
``corrected'' formulae for $\sqrt{s_{NN}}$ = 17.3 GeV for different
values of impact parameter $b$ = 2 (solid), 4 (dashed), 6 (dotted), 
and 8 (dash-dotted) fm. 
}
\label{fig:epsilons_improved}
\end{figure}

Now, when calculating eccentricity coefficients, one could replace $x$
by $x - x_{cm}(\eta_s;b)$ in Eq.(\ref{epsilon_n_definition}).
The so-defined eccentricities are
shown in Fig.\ref{fig:epsilons_improved} as a function of space-time
rapidity.
We observe that the even eccentricity coeeficients are 
relatively large, while odd eccentricity coefficients are rather small 
which is a consequence of the specific, in our opinion
realistic, model considered here.
We observe large reduction of odd eccentricity coefficients compared
to the naive formula (\ref{epsilon_n_definition}).
The modulus of even coefficients grow with rapidity which is a bit different
than in other models in the literature.
It would be pedagogical to compare eccentricity coefficient as
calculated here to other models used in the literature but we do not
feel it is our task at present.
Most of the models describe the elliptic flow coefficients.
The directed flow was much less often studied,
see e.g. \cite{BW2010,ORSMK2020,SA2020}.

The initial eccentricity coefficients $\epsilon_n$'s are transformed
to finally experimentally observed flow coefficient $v_n$'s in 
the preequilibrium \cite{LSH2015} and hydrodynamical \cite{SJG2010,Karpenko}
evolution phases to be compare to experimental data \cite{NA49_vk,STAR}.
This part is to complicated to be discussed here where we wish to 
concentrate exclusively on the initial conditions consistent with local
energy-momentum conservation and baryon number conservation.
Relatively simple relation was found for 
$v_2 \approx 0.25 \epsilon_2$ at midrapidity \cite{NYGO2016} in some
simplified version of hydrodynamical evolution.
Similar sudies could be done for different eccentricity coefficients
from our model in a broad range of rapidity of pions using a version 
of hydrodynamical evolution dedicated to the ``low''-energy scattering 
where plasma is not flavour symmetric and $\mu_B \ne$ 0.
Not all available codes are able to correctly handle such
situations, especially those for our specific model.

\section{Conclusions}

We have studied consequences of local transverse position
(x,y)-dependent energy-momentum conservation for initial 
eccentricities of the plasma created in the noncentral collisions 
of symmetric nuclei (Au+Au or Pb+Pb). These studies are of interest for
collisions performed by the NA51 experiment at SPS and low-energy scan
project at the RHIC collider. 

The following general picture of transformation of initial kinetic energy 
to heat stands behind our model. The first stage of 
the thermalization happens in the longitudinal direction which 
is related to a fast partial stopping of two
pieces of relativistic nuclear matter. Only then the transverse
expansion may start to develop, after some dissipation of initial
kinetic energy.
We wish to point out here that our dynamical fire-streak model 
formulated in \cite{SKR2017} naturally explains torqued initial 
conditions \cite{BBM2011} that lead to forward-backward azimuthal 
correlations.

The proposed approach leads to initial eccentricities of the plasma.
The eccentricity coefficients strongly depend on space-time rapidity 
of the plasma especially that for $\epsilon_1$ or $\epsilon_3$. 
The initial eccentricities are potentially important ingredients 
for final particle flow generated at the freeze-out space-time moment. 
Here we have concentrated on initial conditions and do not consider 
hydrodynamical evolution with such conditions. 
The specialized groups having codes dedicated to
adequate hydrodynamical evolution are welcome to continue the studies.

We have shown that naive calculation of eccentricities with our
initial conditions leads to nophysical results
of eccentricity parameters as a function of $\eta_s$. We have proposed 
a method how to improve the calculation the eccentricity coefficients 
in the case of $(x,y)$-dependent longitudial flow being a consequence of 
initial collision geometry.

We have calculated $\epsilon_{1.2.3,4}$ as a function of space-time
rapidity for different initial impact parameters.
The results strongly depend on the impact parameter as in the earlier
studies with different initial conditions.
In our model of initial longitudinal dynamics the odd eccentricity
coefficients strongly depend on the space-time rapidity.
We wish to note large $\epsilon_2$ and $\epsilon_4$ everywhere and
nonnegligible $\epsilon_1$ and $\epsilon_3$ for large space-time rapidities.
The larger the space-time rapidity the larger the $\epsilon_1$ 
eccentricity parameter. The even initial eccentricity coefficients, 
($\epsilon_2$, $\epsilon_4$), are large, different than 
in other models of initial conditions.
They are rather weakly dependent on space-time rapidity.
Our initial conditions generate also triangular eccentricities
which are usually attributed to event-by-event fluctuations and 
not to initial geometry. The large eccentricity coefficients in
our model are a direct consequence of the energy-momentum conservation
broken in many other approaches in the literature.

We have analysed the initial shapes of the matter created right after
the (peripheral) collisions.
We have shown that the shapes in $(x,y)$ for more peripheral collisions
are rather exotic which is reflected in large eccentricity parameters. 
In our opinion those shapes must evolve very fast already 
in the preequilibrium phase.
The exotic shapes with large gradient of velocity in $z$-direction
cause that the fluctuations in the preequilibrium stage are probably 
crucial but very difficult to put into a mathematical formulation.


\vspace{1cm}

{\bf Acknowledgments}

I am indebted to  Piotr Bo\.zek, Wojciech Broniowski, Jacek Oko{l}owicz
Vitalii Ozvenchouk and Andrzej Rybicki
for several discussions on different aspects of initial conditions 
in relativistic heavy-ion collisions.
The exchange of information with Iurii Karpenko is acknowledged.


\end{document}